\documentclass[aps,prb,reprint,floatfix,superscriptaddress]{revtex4-1}
\usepackage[english]{babel}                                                           
\usepackage[utf8]{inputenc}                                                            
\usepackage[T1]{fontenc}
\usepackage{epsfig}
\usepackage{amsmath}
\usepackage{hyphenat}
\usepackage{hyperref}
\usepackage{placeins}
\usepackage[justification=raggedright]{caption}	
\usepackage[colorinlistoftodos]{todonotes}			  				

\usepackage{array}
\newcolumntype{C}[1]{>{\centering\arraybackslash}m{#1}}
\usepackage{overpic}

\begin{document}

\title{The antiferromagnetic Ising model on the swedenborgite lattice}
\author{Stefan Buhrandt}
\affiliation{Institute for Theoretical Physics and Center for Extreme Matter and Emergent Phenomena,
Utrecht University, Leuvenlaan 4, 3584 CE Utrecht, The Netherlands}
\affiliation{Institut für theoretische Physik, Universität zu Köln, Zülpicher Straße 77, 50937 Köln, Germany}
\email[To whom correspondence should be addressed: ]{buhrandt@thp.uni-koeln.de}
\author{Lars Fritz}
\affiliation{Institute for Theoretical Physics and Center for Extreme Matter and Emergent Phenomena,
Utrecht University, Leuvenlaan 4, 3584 CE Utrecht, The Netherlands}

\begin{abstract}
Geometrical frustration in spin systems often results in a large number of degenerate ground states. In this work we study the antiferromagnetic Ising model on the three dimensional swedenborgite lattice which is a specific stacking of Kagom\'e and triangular layers. The model contains two exchange couplings, one within the Kagom\'e layer, another one in between Kagom\'e and triangular layers. We determine the phase diagram with and without easy axis magnetic field and calculate the ground state degeneracy explicitly in terms of the residual entropy. At zero field we find two different ground state manifolds separated by a first order transition at $T=0$ and equal exchange couplings. We also determine the $T=0$ phase diagram in a magnetic field and find a rich phase diagram with both degenerate and non-degenerate phases depending on the field strength and out-of-plane coupling.
\end{abstract}
\maketitle

\section{Introduction}

\begin{figure}
	\centering
	\includegraphics[width=\columnwidth]{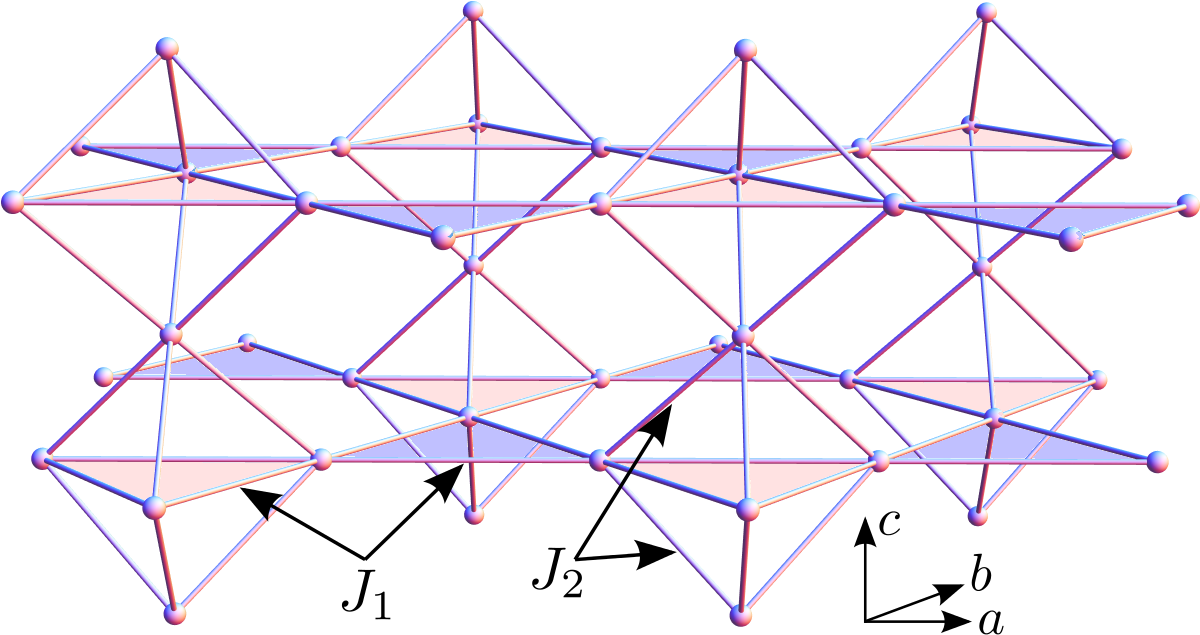}
	\caption{The lattice formed by the magnetic ions in swedenborgite compounds. The elementary units considered in this paper are the bipyramids which are joined via intermediate triangles (light blue).}
	\label{fig:lattice}
\end{figure}

Frustration can exist when conflicting interactions cannot mutually be satisfied. A prime example are antiferromagnetically coupled spins on geometrically frustrated lattices. Geometric frustration, for instance present in two dimensions (2d) on the triangular or Kagomé lattices or in three dimensions (3d) on the pyrochlore lattice, does not automatically entail the existence of degenerate classical ground states, the hallmark of frustration. Whether degeneracies occur depends on the one hand on the coordination number of the lattice and on the other hand on the symmetry of the spins. While classical $O(3)$ Heisenberg spins on the triangular lattice have a unique ground state with $120^{\circ}$ order, there is a large number of degenerate ground states on the Kagom\'e lattice \cite{Chalker1992}. Discrete Ising spins on the other hand result in an extensive ground state degeneracy on both lattices. In many cases, this degeneracy can either be reduced or lifted completely by applying an external magnetic field.\\
Recently, a new class of geometrically frustrated structures based on cobalt oxides, RBaCo$_4$O$_7$ , where R denotes a rare earth atom, emerged~\cite{Valldor2009,Chapon2006,Soda2006,Valldor2002,Manuel2009,Khalyavin2010,Schweika2007,Valldor2006}. The magnetic Co-ions in this structure reside on the so-called swedenborgite lattice shown in Fig.~\ref{fig:lattice}. Consisting of triangle sharing bipyramids, it has a rather unique exchange topology and offers a perfect playground to study the effect of geometric frustration in detail. We recently argued that $O(3)$ Heisenberg spins on this lattice exhibit a huge spin liquid regime for a certain parameter and temperature range and can undergo an order-by-disorder transition to a nematic phase at very low temperatures~\cite{Buhrandt2014}. Here, we analyze the Ising model with and without magnetic field on this lattice and quantify the ground state degeneracy in terms of the residual entropy at $T=0$.

The organization of the paper is as follows: we first introduce the model as well as our theoretical approach in Sec.~\ref{sec:model}. In Sec.~\ref{sec:pdb0} we determine the $T=0$ and $B=0$ phase diagram and the associated ground state degeneracies. In Sec.~\ref{sec:pdbf} we repeat this analysis for finite magnetic field and we finish with concluding remarks in Sec.~\ref{sec:conclusion}.

\section{Model and approach}\label{sec:model}

The swedenborgite lattice has a 3d hexagonal lattice structure with a non-trivial unit cell comprised out of eight atoms, see Fig.~\ref{fig:lattice}. This lattice can either be understood as an alternating stack of Kagom\'e and triangular layers in an ABAC... pattern (A=triangular layer, B,C = non-equivalent Kagom\'e layer) or equivalently as bipyramidal clusters connected by intermediate triangles in the ab-plane and stacked along c-direction. The latter viewpoint makes the unique exchange topology of the lattice more visible. Whereas most other frustrated lattices are constructed of \emph{corner}-sharing frustrated units like for instance triangles on the Kagom\'e- or tetrahedra on the pyrochlore lattice, the swedenborgite lattice is constructed out of \emph{triangle}-sharing bipyramids, \textit{i.\,e.} frustrated clusters which share other frustrated clusters with each other.\\  
We consider a nearest neighbor ($\langle i,j \rangle$) Ising model with only two distinct antiferromagnetic interactions: $J_1$ inside the Kagom\'e layers and $J_2$ between the Kagom\'e and triangular layers,
\begin{equation}
	H = J_1\sum_{ {\langle i,j \rangle \in \atop \text{same layer}}} \sigma_i \sigma_j + J_2\sum_{{\langle i,j \rangle \in \atop \text{diff. layer}}} \sigma_i \sigma_j - B \sum_i \sigma_i,
	\label{eq:Hamiltonian}
\end{equation}
with Ising spins $\sigma_i = \pm 1$ and magnetic field $B$ pointing along the easy spin axis. A magnetic field with other components would lead to a quantum model which will be subject of a forthcoming publication. \\

We analyze this model by a combination of mean-field theory and classical Monte-Carlo simulations. The former one is used to obtain the $T=0$ phase diagram of the model with and without magnetic field. Based on this analysis, we will present arguments why the ground state of the model is either extensive, sub-extensive or not degenerate at all. In order to compare and quantify the degree of degeneracy for different parameter values we compute the residual entropy $S_{\text{res}}$ of the ground state at $T=0$. Integrating the thermodynamic relation $dS/dT = c_V/T$ from $T=0$ to $T=\infty$ and using the exact known result for the entropy of a free Ising spin, $S(T=\infty)=\ln 2$, we obtain 
\begin{equation}
	S_{\text{res}} = \ln 2 - \int_{0}^{\infty} \frac{c_V}{T} dT \;.
	\label{eq:residual_entropy}
\end{equation}
The specific heat entering this equation is determined numerically with classical Monte Carlo simulations by the fluctuations of the energy,
\begin{equation}
	c_V = \frac{\langle E^2 \rangle - \langle E \rangle^2}{NT^2}\;.
\end{equation}
To obtain accurate results for the residual entropy, it is necessary to calculate the specific heat with high accuracy. We found that simple simulated annealing algorithms can easily become non-ergodic if the out-of-plane coupling $J_2$ is either weak or about the the same size as the in-plane coupling $J_1$. It turns out that this problem can be solved using parallel tempering Monte Carlo together with a feedback algorithm \cite{Katzgraber2006} that chooses temperature points such that the current of replicas drifting through temperature is maximized. This typically results in a dense grid of temperature points close to the phase transition or freezing temperature. Starting from an initial temperature set following a geometric progression, convergence of the feedback algorithm was typically reached after one to three iterations. Using the optimized temperature set we found that we could ensure ergodicity for all parameter values used in our simulations by monitoring the replica current in temperature space. Simulations were performed on a lattice with $N=8L^3$ lattice sites and $L=\{6,12,18\}$.

\section{Phase diagram and ground state degeneracy for $\mathbf{B=0}$}\label{sec:pdb0}

\begin{table}
	\centering
	\begin{tabular}{C{1.5cm}|C{6cm}}
		            &	allowed bipyramid configurations \\ \hline 
		$J_2/J_1>1$ &	\includegraphics[width=1.1cm]{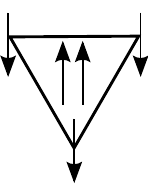} \\
		$J_2/J_1=1$ &	\includegraphics[width=1.1cm]{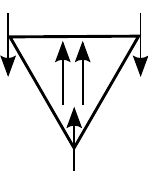}\quad\includegraphics[width=1.1cm]{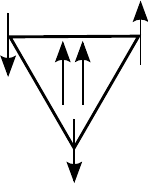}\quad\includegraphics[width=1.1cm]{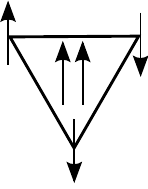}\quad\includegraphics[width=1.1cm]{BP1} \\
		$J_2/J_1<1$ &	\includegraphics[width=1.1cm]{BP2}\quad\includegraphics[width=1.1cm]{BP3}\quad\includegraphics[width=1.1cm]{BP4} \\
	\end{tabular}
	\caption{The allowed bipyramid configurations in the different ground state regimes. The two spins in the middle of each triangle denote the apical spins of a bipyramid which are assumed to be already fixed in ``up'' direction. For a 3d figure of the bipyramid configurations, see insets of Fig.~\ref{fig:residual_entropies}. The configurations for $J_2/J_1>1$ and $J_2/J_1<1$ have an energy per unit cell of $E=-12J_2+6J_1$ and $E=-8J_2+2J_1$, respectively.}
	\label{tab:allBP}
\end{table}
We start with a discussion of the elementary building blocks of our system. The intermediate triangles without apical spins on top and underneath (see blue triangles in Fig.~\ref{fig:lattice}) always take an up-up-down or up-down-down configuration independent of $J_1$ and are six-fold degenerate, just like in a standard triangular lattice~\cite{Wannier1950}.\\
Within the bipyramids the situation is slightly more complicated since the ground state configuration depends on the ratio $J_2/J_1$, see Tab.~\ref{tab:allBP} for an overview. For $J_2/J_1>1$ the three Kagom\'e spins are aligned mutually parallel and anti-parallel to the apical spins (first row in Tab.~\ref{tab:allBP}). This configuration has energy $E=-12J_2+6J_1$ per unit cell and is not degenerate if one assumes that the apical spins are fixed. For $J_2/J_1<1$, it becomes favorable for the system to flip one of the three Kagom\'e spins in each bipyramid to align it parallel with the apical spins (third row in Tab.~\ref{tab:allBP}). In this configuration with energy $E=-8J_2+2J_1$ per unit cell each bipyramid is three-fold degenerate. For $J_2/J_1=1$, both above mentioned bipyramid configurations have the same energy and the bipyramids are thus four-fold degenerate.\\
The ground state manifold of the whole lattice is constructed by connecting the respective bipyramids either via up-up-down or up-down-down triangles within the plane. The system possesses a large degeneracy in the ab-plane due to the intermediate triangles irrespective of the ratio $J_2/J_1$, corresponding to extensive ground state entropy with respect to the ab-plane. However, for the full three dimensional system the degree of degeneracy depends on the c-direction. For $J_2/J_1>1$, the bipyramid configuration in one plane fixes all bipyramids in c-direction since once the configuration of the apical spins within a column (stack of bipyramids on top of each other in c-direction) is fixed there is no freedom left, meaning the ground state entropy is subextensive.\\
For $J_2/J_1<1$ the direction of the apical spins does not fix entire columns any more but bipyramids on top of each other can choose Kagom\'e spin configurations independently of each other. This implies that fixing all bipyramids in one plane still allows to choose the spin configurations in the other planes independently (except for the apical spins). The system is thus extensively degenerate in this case with a finite residual entropy in the thermodynamic limit.\\
At the equal coupling point, $J_2/J_1=1$, the ground state degeneracy is extensive and even larger than for $J_2/J_1<1$ due to the 4-fold degenerate bipyramids the system can choose from.\\ 

\begin{figure}
	\centering
	\begin{overpic}[]{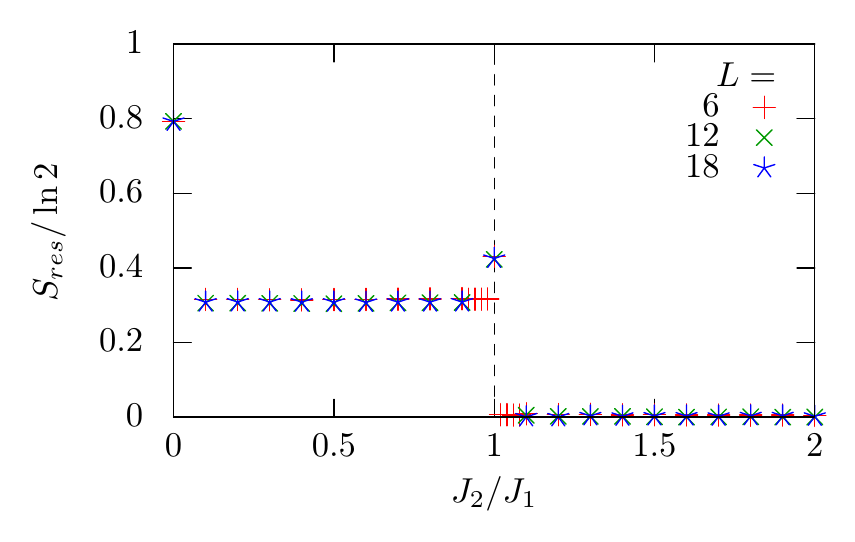}
		\put(26,32){\includegraphics[width=2.4cm]{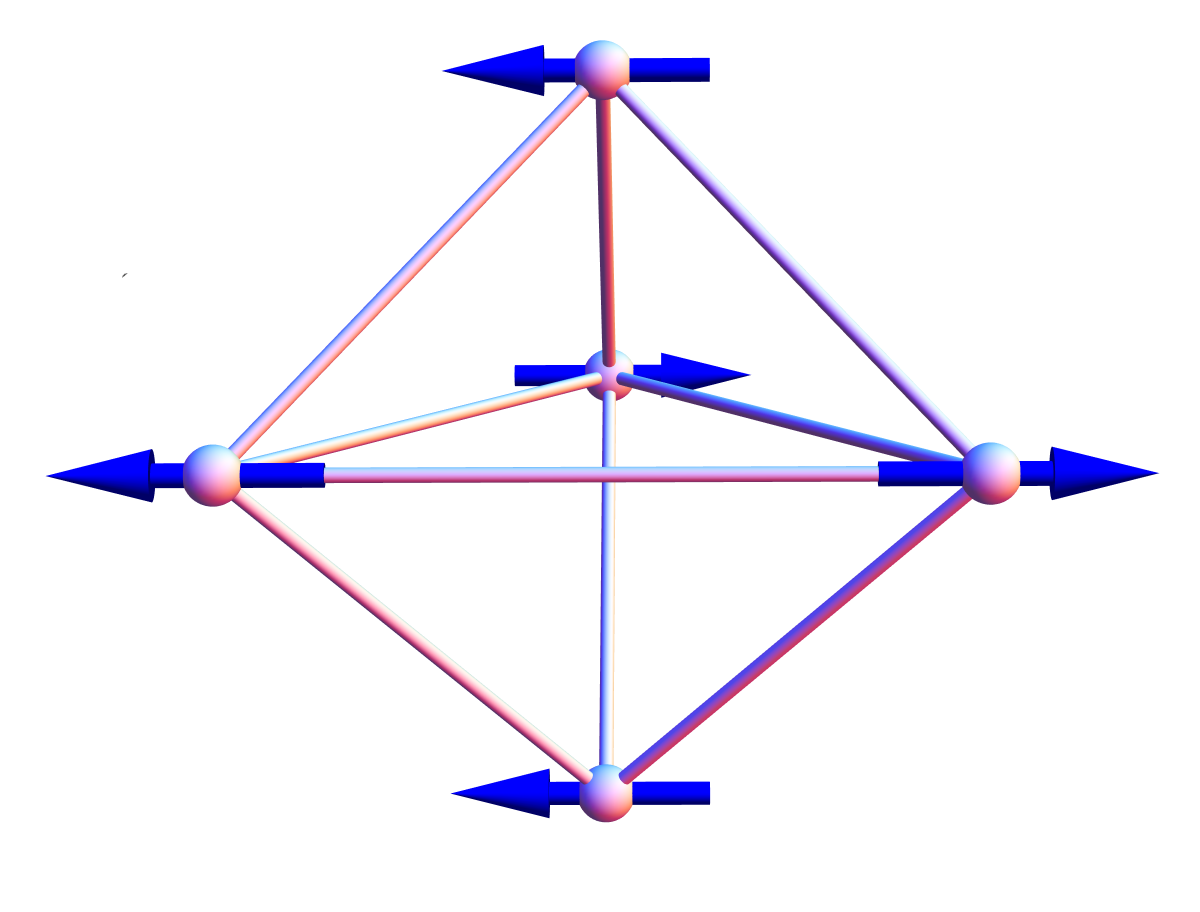}}
		\put(63,18){\includegraphics[width=2.4cm]{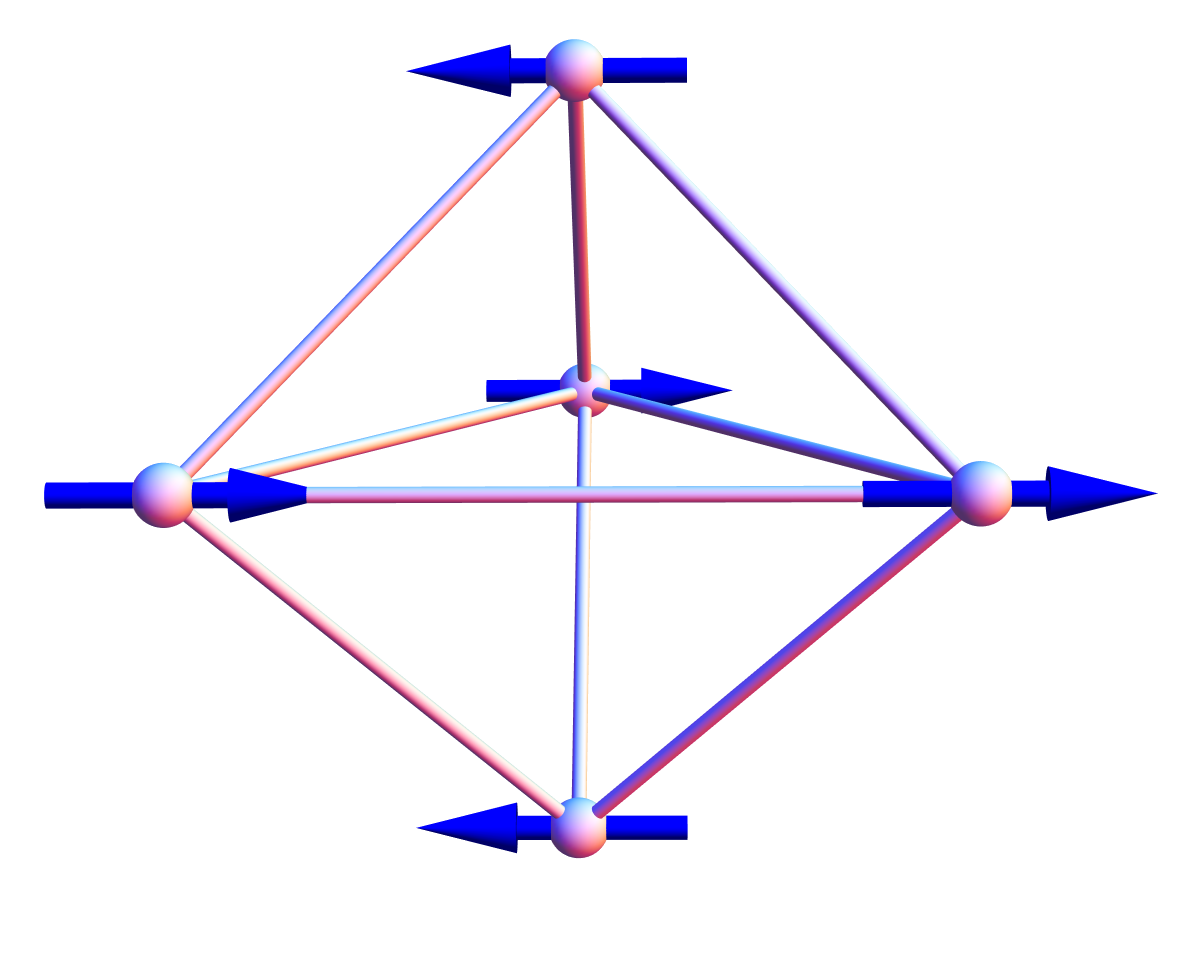}}
	\end{overpic}
	\caption{Residual entropies for the Ising model without magnetic field calculated with Eq.~\eqref{eq:residual_entropy}. For $J_2/J_1<1$ the ground state is extensively degenerate whereas it is sub-extensively degenerate for $J_2/J_1>1$. At $J_2/J_1=1$, the degeneracy is enhanced due to the additional degeneracy of the bipyramids. The insets show the spin configurations on the bipyramids for $J_2/J_1<1$ (left) and $J_2/J_1>1$ (right). }
	\label{fig:residual_entropies}
\end{figure}

We have calculated the residual entropy for various values of $J_2/J_1$ in the range from 0 to 2, see Fig.~\ref{fig:residual_entropies}. For $J_2/J_1<1$ we find a value of $S_{\text{res}}/\ln 2 \approx 0.32$ which is clearly smaller than the values reported for the AFM Ising model on the triangular lattice ($S_{\text{res}}/\ln 2 \approx 0.47$, \cite{Loh2008}) and the Kagom\'e ($S_{\text{res}}/\ln 2 \approx 0.72$, \cite{Kano1953,Loh2008}), but larger than for so-called ``spin-ice'', the AFM Ising model on the pyrochlore lattice ($S_{\text{res}}/\ln 2 \approx 0.29$ \cite{Gingras2009}). For equal exchange couplings we find the residual entropy to be about $S_{\text{res}}/\ln 2 \approx 0.43$ and thus about $4/3$ times as large as for $J_2/J_1<1$ as one would naively expect from the enhanced degeneracy of the bipyramids. Our findings are summarized in Tab.~\ref{tab:Sres}.\\
As expected from our analytical analysis of the ground state manifold, the residual entropy is constant in the two ground state regimes for $J_2/J_1<1$ and $J_2/J_1>1$ and does not depend on the actual ratio $J_2/J_1$ in the respective regimes. 

\begin{table}
	\centering
	\begin{tabular}{c|c}
	Lattice	&	$S_{\text{res}}/\ln 2$  \\ \hline \hline
	Triangular	& 0.47	\cite{Loh2008}\\
	Kagom\'e	& 0.72 \cite{Kano1953,Loh2008}	\\
	Pyrochlore & 0.29 \cite{Gingras2009}\\
	Swedenborgite ($J_2/J_1<1$) &	0.32\\
	Swedenborgite ($J_2/J_1=1$) & 0.43	\\
	\end{tabular}		
	\caption{Residual entropies at $T=0$ and $B=0$ for the AFM Ising model on different lattices.}
	\label{tab:Sres}
\end{table}

\begin{figure}
	\centering
	\includegraphics{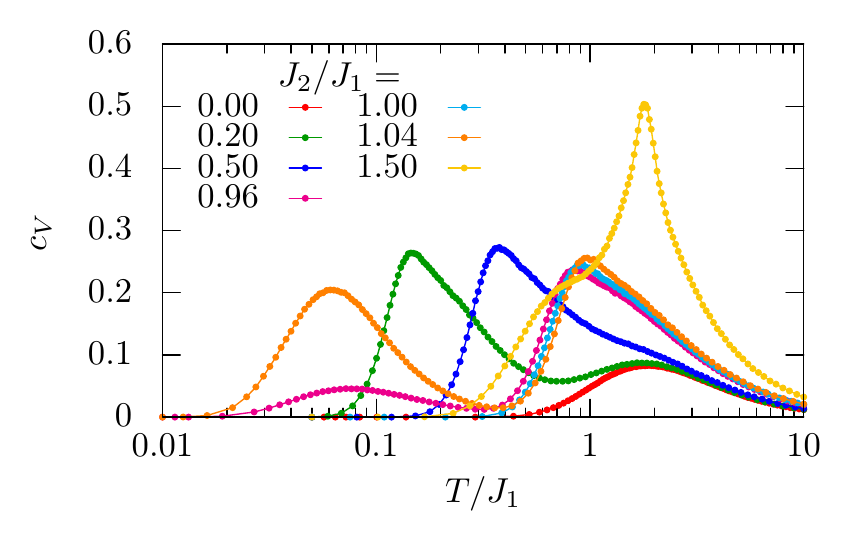}
	\caption{Specific heat of the Ising model with $B=0$, system size $L=6 (N=1728)$ and various ratios of $J_2/J_1$. If $J_2$ is small or close to $J_1$, the freezing temperature of the system is shifted to very low temperatures, see main text.}
	\label{fig:cv_ising}
\end{figure}

\begin{figure}
	\centering
	\begin{overpic}[]{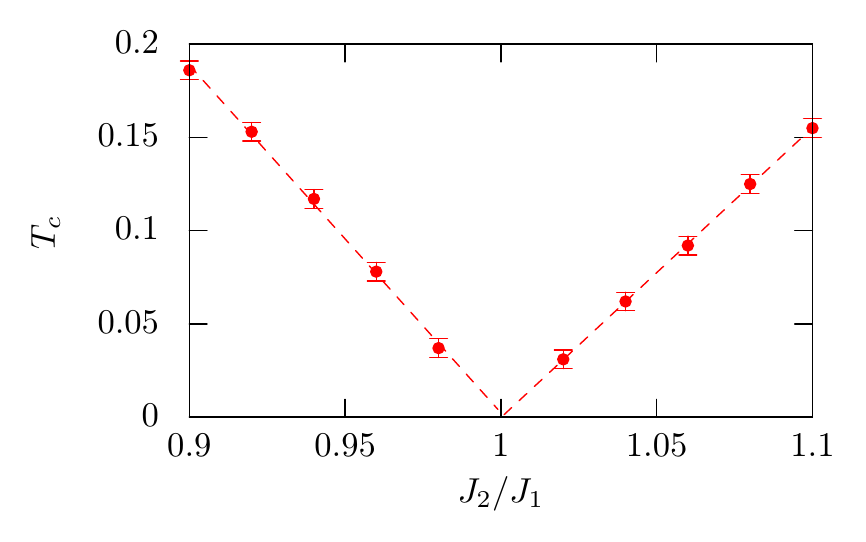}
		\put(25,15){\includegraphics[width=1.9cm]{Bipyramid_2}}
		\put(72,15){\includegraphics[width=1.9cm]{Bipyramid_1}}
		\put(36,35){\includegraphics[width=1.9cm]{Bipyramid_2}}
		\put(62,34.5){\includegraphics[width=1.9cm]{Bipyramid_1}}
		\put(57.5,42){\large{\textbf{+}}}
		\put(50,26){$1^{\text{st}}$ order}
		\put(57.9,25){\vector(0,-1){8}}
	\end{overpic}
	\caption{The crossover temperature to the different ground states regimes determined from the position of the low temperature hump in the specific heat as a function of $J_2/J_1$. The dashed lines represent a linear regression.}
	\label{fig:QCP}
\end{figure}

Whenever the system locks into a specific ground state manifold, its entropy decreases. Since $c_V = T dS/dT$, this decrease of the entropy is accompanied by a hump in the specific heat, see Fig.~\ref{fig:cv_ising}. If the out-of-plane coupling is much weaker than the in-plane coupling, the system crosses over into its ground state manifold in two steps. First, the intermediate triangles select up-up-down and up-down-down states, resulting in a hump at a temperature of order $\mathcal{O}(J_1)$, whereas the bipyramids remain disordered until a much lower temperature of order $\mathcal{O}(J_2)$ where they eventually enter their respective ground state configurations, accompanied by a second hump in the specific heat. These two peaks merge into one broad peak as $J_2/J_1$ is increased above $\approx 0.3$ and the two crossovers cannot be separated any more.\\
Another interesting parameter region is given by almost equal exchange couplings. Exactly at $J_2/J_1=1$, we find only one hump at a temperature of $\mathcal{O}(J_1)$, indicating that the intermediate triangles and the bipyramids enter their respective ground state manifold at the same time. In the vicinity of this point, {\it i.e.} $J_2=J_1\pm \delta$, we observe an additional well separated hump at low temperatures which is pushed towards $T=0$ as $\delta \to 0$. This feature originates from the fact that the two relevant bipyramid configurations, which are truly degenerate at $J_2=J_1$, appear \emph{almost} degenerate for $J_2\approx J_1$ until the temperature becomes of the order of their energy splitting $\Delta E = |J_2-J_1|/2$ per spin. As a consequence, the position of the associated hump in the specific heat is found to approach $T=0$ linearly as $J_2/J_1$ approaches $1$, see Fig.~\ref{fig:QCP}. Exactly at $J_2/J_1=1$ and $T=0$ there is a first order phase transition between the two ground state regimes due to a level crossing of the two ground state configurations of the bipyramids, see Tab.~\ref{tab:allBP}.\\

\subsection{Structure factors}

\begin{figure*}
	\includegraphics[width=\textwidth]{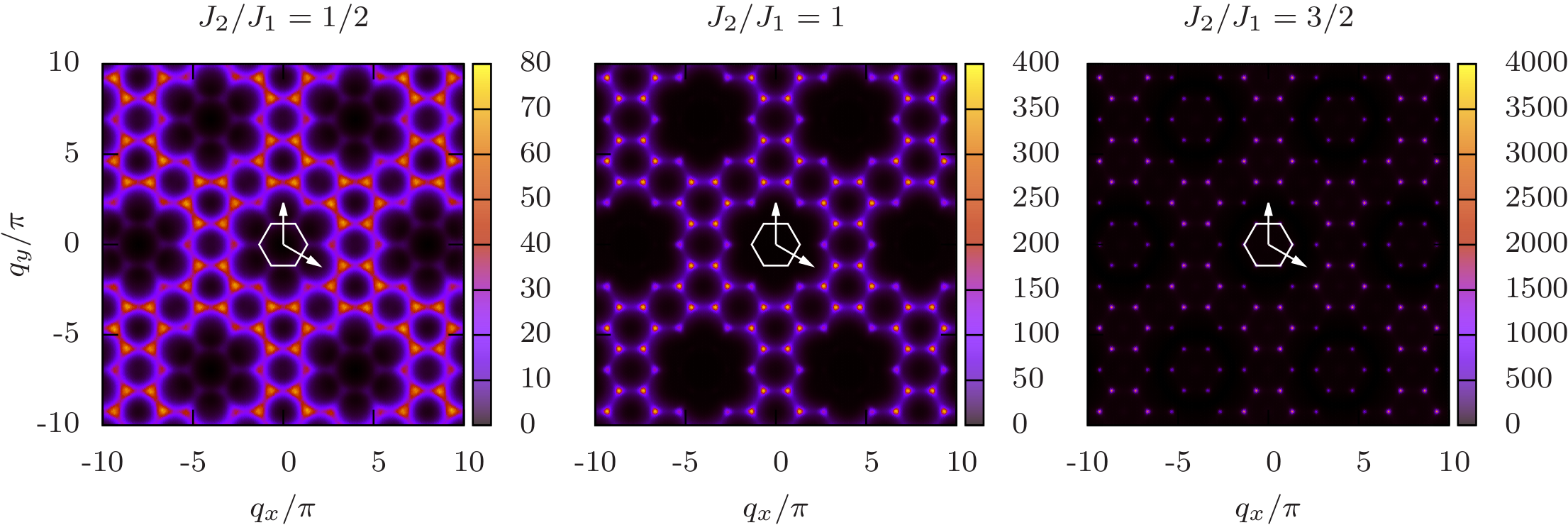}
	\caption{The magnetic structure factor Eq.~\eqref{eq:sf} in the $(q_x, q_y, q_z=0)$ plane at $T=0$, $B=0$ and $L=30$. The white arrows denote the reciprocal lattice vectors and the white hexagon represents the first Brillouin zone. One can clearly observe a broadening of the peaks for $J_2/J_1 \leq 1$.}
	\label{fig:sf}
\end{figure*}

\begin{figure}
	\centering
	\includegraphics{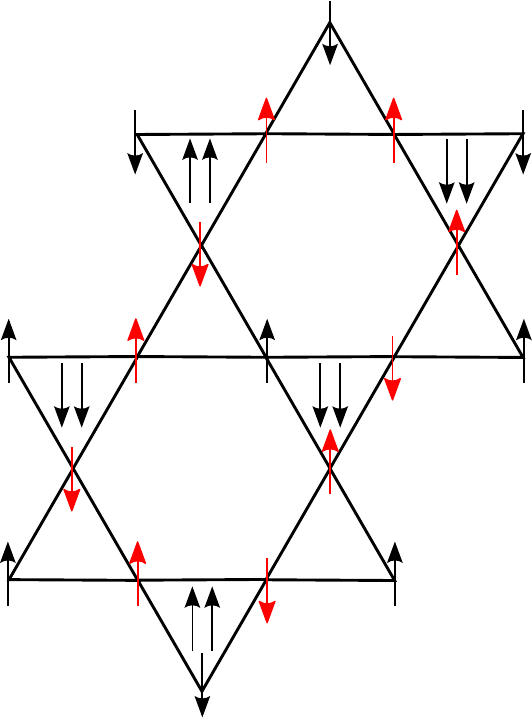}
	\caption{A zero-energy loop spin flip in the ground state manifold for $J_2/J_1<1$. Flipping the chain of red spins at the same time does not cost any energy and the system remains in the ground state manifold.}
	\label{fig:loopflip}
\end{figure}

We have also calculated the magnetic structure factor 
\begin{equation}
	S(\mathbf{q}) =\frac{1}{N}\sum_{ij} \langle \mathbf{S}_i \cdot \mathbf{S}_j \rangle e^{i \mathbf{q}  (\mathbf{r}_i - \mathbf{r}_j)} 
	\label{eq:sf}
\end{equation}
for $J_2/J_1=1$, $J_2/J_1<1$ and $J_2/J_1>1$, see Fig.~\ref{fig:sf}. Again, the precise value of $J_2/J_1$ has no influence on the result in the latter two regimes. One can clearly see that the structure factor shows stronger correlations as $J_2/J_1$ is increased above 1. Responsible for this are the correlations along c-direction: For $J_2/J_1>1$, all Kagom\'e spins of a stacked bipyramid cluster are anti-parallel to the triangular spins of the same cluster, while this is only the case for 3/4 and 2/3 of the spins for $J_2/J_1=1$ and $J_2/J_1<1$, respectively. For $J_2/J_1<1$, the structure factor shows features reminiscent of pinch points which are well known to occur in frustrated spin systems governed by local constraints~\cite{Henley2009,Zhitomirsky2008}: In these systems, it is often possible to describe the ground state manifold as a so-called Coulomb phase, where a divergence free field $\mathbf{E}(\mathbf{r})$ is associated with each vertex of the lattice in the ground state (see e.\,g. Ref.~\onlinecite{Henley2009} for a review). In these models, excitations above the ground state appear as pairwise local charge defects which interact via an effective Coulomb potential $V(\mathbf{r}) \propto 1/|\mathbf{r}|$. One direct consequence of the divergence-free constraint is that correlations decay as
\begin{equation}
	\langle E^{\mu}(-\mathbf{q}) E^{\nu}(\mathbf{q}') \rangle  \propto  \left(\delta_{\mu\nu} - \frac{q_{\mu}q_{\nu}}{\mathbf{q}^2} \right) \delta_{\mathbf{q},\mathbf{q'}} 
	\label{eq:pinch}
\end{equation}
in momentum space, leading to the above mentioned pinch-points in the structure factor.\\
The requirements for the existence of such a Coulomb phase are that the corresponding medial lattice is bipartite and that it is possible to describe the ground state manifold either by an ``ice-rule'' or a hardcore dimer covering on this lattice \cite{Henley2009}. An ice-rule description can only be found if the medial lattice has an even coordination number; the spins on the original lattice define a flux on the medial lattice and the sum of all fluxes into a node on the latter vanishes for every ground state spin configuration. Mappings to hardcore dimer models can be found only in some model specific cases. The fact that the bipyramids and triangles on the swedenborgite lattice consist of an odd number of spins excludes an ice-rule description for our model and we also believe that no mapping of the ground state manifold to a dimer covering exists. A detailed analysis of the region around the bow-tie like features observed in the structure factor for $J_2/J_1<1$ reveals that there is no direction in which the correlation vanishes completely as one moves away from the center of the bow-tie (as it should for a real pinch point according to Eq.~\eqref{eq:pinch}). Since the structure factor was obtained strictly at $T=0$, thermal broadening of a true pinch point cannot be the reason for this observation. The observed bow-tie like features thus do not seem to be pinch points originating from a Coulomb phase of our model.\\
There are nevertheless some similarities between a Coulomb phase and the ground state manifold for $J_2/J_1<1$. Coulomb phases are well known to host loops of zero energy spin flips which connect different ground states with each other. A non-closed loop also allows charge defects to move without energy cost through the system and is thus the origin of the defect mobility in these phases. We find similar loops within the Kagom\'e layers of the swedenborgite lattice, corresponding to flipping a loop of neighboring anti-ferromagnetically aligned spins, see Fig~\ref{fig:loopflip}. In contrast to Coulomb phases, it is however not possible to reach every ground state with these loop moves since the spins on the triangular lattice remain unaffected. Nevertheless, these chains mediate the movement of single defects, in lowest order described by bipyramids with three instead of two Kagom\'e spins antiparallel to the apical ones, within the Kagom\'e planes.\\
One can also easily construct such zero energy spin flips for $J_2/J_1=1$, where the loops do not necessarily have to be closed due to the enhanced degeneracy of the ground state manifold, whereas no such moves exist for $J_2/J_1>1$.\\ 
    
\section{Phase diagram and ground state degeneracy for $\mathbf{B \neq 0}$}\label{sec:pdbf}

\begin{figure}
	\centering
	\includegraphics{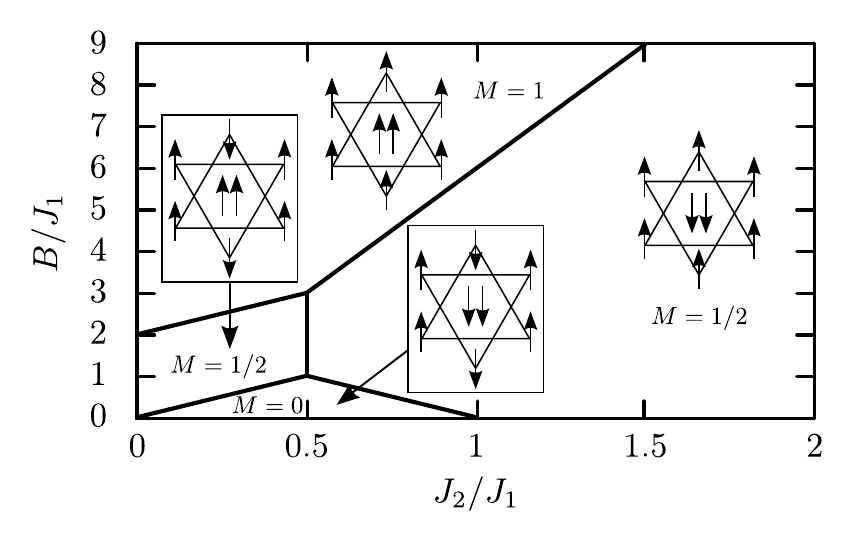}
	\caption{$T=0$ phase diagram for the stacked bipyramids shown in Tab.~\ref{tab:BPconfigurations} in a magnetic field. $M$ denotes the magnetization per spin in the respective phases. All transitions between the different states are first order.}
	\label{fig:phase_diagram_columns}
\end{figure}

The ground state degeneracy that we encountered in the previous section can be lifted partially or totally by a magnetic field. In this section we study this process in detail and present the T = 0 phase diagram which features a large variety of phases with different degeneracies and magnetizations.\\
On the intermediate triangles, an infinitesimal magnetic field already favors the up-up-down configurations and thus reduces the degeneracy by a factor of two to three-fold. This configuration with energy $E=-J_1-B$ per triangle becomes unfavorable with respect to the (non-degenerate) fully polarized state with energy $E =3J_1-3B$ at $B/J_1=2$.\\
There are in total six configurations of stacked bipyramids which are either unaffected by a magnetic field or can gain energy from it. These configurations are shown in Tab.~\ref{tab:BPconfigurations} with the corresponding $T=0$ phase diagram in Fig.~\ref{fig:phase_diagram_columns}.
\begin{table}[h]
\centering
	\begin{tabular}{C{1.6cm}|C{2.1cm}||C{1.6cm}|C{2.1cm}}
		state & energy  & state & energy \\ \hline  \hline
		\includegraphics[scale=1.0]{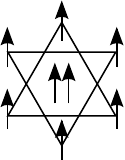} &  $12J_2+6J_1 \newline -8B$ &
		\includegraphics[scale=1.0]{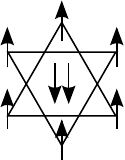} &  $-12J_2+6J_1 \newline -4B$ \\
		\includegraphics[scale=1.0]{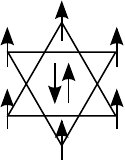} &  $6J_1-6B$ &
		\includegraphics[scale=1.0]{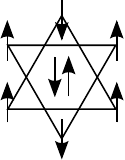} &  $-2J_1 -2B$ \\
		\includegraphics[scale=1.0]{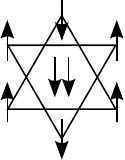} &  $-4J_2-2J_1$ &
		\includegraphics[scale=1.0]{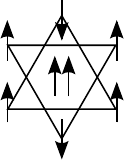} &  $4J_2-2J_1 \newline -4B$ \\
	\end{tabular}
	\caption{List of all stacked bipyramid configurations that are either unaffected by a magnetic field or gain energy from it together with their respective energy per unit cell.}
	\label{tab:BPconfigurations}
\end{table}

\begin{figure}
	\centering
	\includegraphics{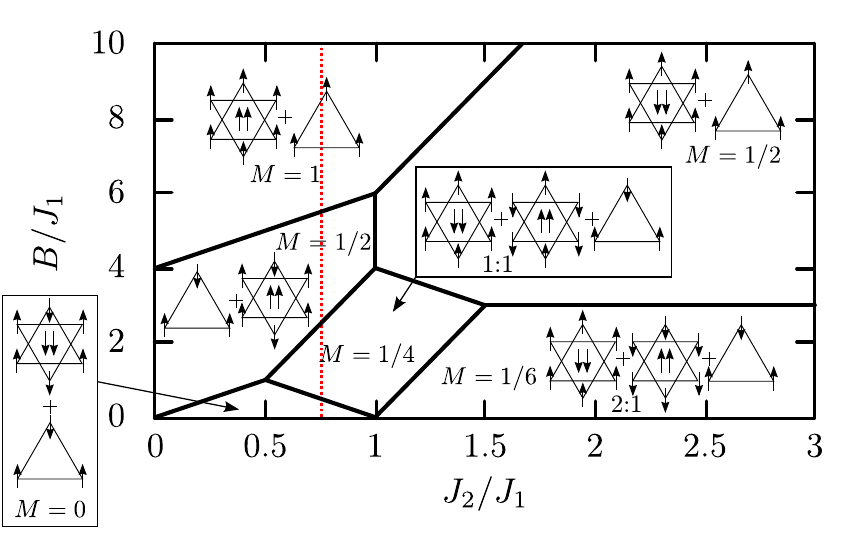}
	\caption{$T=0$ Phase diagram of the Ising model on the swedenborgite lattice in an external magnetic field. ``up'' pointing triangles represent the blue intermediate triangles from Fig.~\ref{fig:lattice} and ``down'' pointing triangles represent the triangles inside the bipyramids (red triangles in Fig.~\ref{fig:lattice}) with the two apical spins shown inside. All transitions between the different phases are first order. Along the red line the residual entropy changes as $0.24\ln2 \to 0 \to 0.11\ln2 \to  0$ in the different phases as the magnetic field is increased.}
	\label{fig:phase_diagram_B_vs_J2}
\end{figure}

While the determination of the phase diagram for isolated triangles and stacked bipyramidal clusters is rather straightforward, the situation becomes much more complicated when these units are connected as on the swedenborgite lattice. There are certain stacked bipyramid configurations (\textit{e. g.} the first configuration in Tab.~\ref{tab:BPconfigurations}) that cannot be connected to an intermediate up-up-down triangle without an energy penalty due to the magnetic field even though the single bipyramid might be able to gain energy from the field if rotated in the right direction. On the other hand, it might be favorable to have up-up-down configurations on the intermediate triangles even though an isolated triangle would prefer to be fully polarized. The only way to find the phase diagram is to systematically write down all possible bipyramid configurations and combine them with intermediate up-up-down or up-up-up triangles. This task becomes even more complicated if one takes into account that not all bipyramids in the system need to have the same configuration.\\
We have compared the energy of all possible combinations and checked our results against classical Monte Carlo simulations. The resulting $T=0$ phase diagram is shown in Fig.~\ref{fig:phase_diagram_B_vs_J2}.\\
We find that there are in total six phases at $T=0$, four of which differ in their magnetization. The transitions between these phases are again all first order as they originate from level crossings of the respective energies. We have also calculated the residual entropies and find that only the $M=0$ phase ($S_{\text{res}}/\ln 2 \approx 0.24$) and the $M=1/2$ phase for $J_2/J_1<1$ ($S_{\text{res}}/\ln 2 \approx 0.11$ have an extensive ground state degeneracy. The difference in the residual entropy of a factor of $\approx 2$ comes from the fact that in the $M=0$ phase the bipyramids can be rotated by 180$^{\circ}$ around the c-axis without changing their energy whereas this is not possible in the $M=1/2$ phase for $J_2/J_1<1$.\\
In general, the ground state degeneracy is always smaller than for zero field, as expected. There are, however, regions in the phase diagram where the degeneracy changes unexpectedly as the field is varied at constant $J_2/J_1$. Following the red line in Fig.~\ref{fig:phase_diagram_B_vs_J2}, the residual entropy changes according to $0.24\ln2 \to 0 \to 0.11\ln2 \to  0$ in the different phases as the magnetic field is increased, \textit{i.\,e.} at some point an increase of the magnetic field counter-intuitively leads to an increase of the degeneracy. The reason is rooted in the fact that the ground state manifold is more rigid in the intermediate $M=1/4$ zero entropy phase than the two adjacent finite entropy phases. The factor of approximately two in the residual entropy between the two phases traces back to the fact that while in the $M=0$ phase the bipyramids are two-fold degenerate, a unique alignment of the bipyramids is favored in the $M=1/2, J_2/J_1<1$ phase. 

\section{Conclusion and Outlook}\label{sec:conclusion}

In this paper we have analyzed the anti-ferromagnetic Ising model on the swedenborgite lattice with and without magnetic field. At zero field, we found two different ground state regimes for $J_2/J_1>1$ and $J_2/J_1<1$ separated by a first order transition at $T=0$. In the vicinity of this point, the crossover to the respective ground state manifold happens in two stages and the temperature of the second crossover is linearly shifted to $T=0$ as $J_2/J_1 \to 1$ due to the small energy difference $\Delta E \propto |J_2-J_1|$ between the two competing ground state regimes. We have further calculated the $T=0$ phase diagram in the $B-J_2$ plane and found a rich phase diagram with six different ground state manifolds. In two of these ground states manifolds, the degeneracy is only partially lifted by the applied field.\\
It would be interesting to study the effect of a transverse magnetic field to the ground state manifold for $J_2/J_1<1$. This field would give rise to a term $\propto -B^x\sum_i \sigma_i^x$ in the Hamiltonian that allows the system to gain energy from spin flips. This might result in a selection of ground states with a maximal number of flippable loops in an order-by-disorder transition at infinitesimal transverse fields. We will study this effect in a forthcoming publication.\\

\section{Acknowledgments}

We acknowledge discussions with K. P. Schmidt, J. Reim, W. Schweika and S. Trebst. This work was supported by the Deutsche Forschungsgemeinschaft within the Emmy-Noether program through Grant No. FR 2627/3-1 (S.B. and L.F.) and the Bonn-Cologne Graduate School for Physics and Astronomy (S.B.). This work is also part of the D-ITP consortium, a program of the Netherlands Organisation for Scientific Research (NWO) that is funded by the Dutch Ministry of Education, Culture and Science (OCW). Simulations were performed on the CHEOPS cluster at RRZK Köln.

\bibliography{references}

\end{document}